\documentstyle[12pt]{article}

\newcommand{\be}{\begin{equation}}
\newcommand{\bea}{\begin{eqnarray}}
\newcommand{\ee}{\end{equation}}
\newcommand{\eea}{\end{eqnarray}}

\def\theequation{\arabic{section}.\arabic{equation}}
\begin{document}
\topmargin -1cm \oddsidemargin=0.25cm\evensidemargin=0.25cm
\setcounter{page}0
\renewcommand{\thefootnote}{\fnsymbol{footnote}}
\begin{titlepage}
\begin{flushright}
hep-th/0607248\\
\end{flushright}
\vskip .7in
\begin{center}
{\Large \bf On Lagrangian Formulation of Higher Spin Theories on
AdS} \vskip .7in {\large Angelos Fotopoulos$^a$ \footnote{e-mail:
{\tt afotopou@physics.uoc.gr}}, Kamal L.
Panigrahi}$^{b}$ \footnote{e-mail:{\tt panigrahi@iitg.ernet.in}}
and {\large Mirian Tsulaia$^a$} \footnote{e-mail: {\tt
tsulaia@physics.uoc.gr}} \vskip .2in
{$^a$\it Department of Physics, University of Crete, 710 03 Heraklion, Crete, Greece} \\
\vskip .2in {$^b$ \it Department of Physics,
Indian Institute of Technology, Guwahati, 781 039 India}\\

\end{center}
\vskip .7in
\begin{abstract}
In this short note we present a Lagrangian formulation for free
bosonic  Higher Spin fields which belong to massless reducible
representations of ${\cal D}$-dimensional Anti de Sitter group
using an ambient space formalism.

\end{abstract}
\vfill
\end{titlepage}
\tableofcontents
\setcounter{equation}0
\section{Introduction}

Massless Higher Spin gauge theories are classical field theories
which describe (self)-interacting massless fields with an
arbitrary spin (see \cite{Vasiliev:2004qz} for recent reviews).
Being an interesting subject by itself Higher Spin gauge theories
recently triggered an increasing interest because of their
possible connection with String and M-theory, namely it has been
conjectured that massless Higher Spin theory is the most
symmetrical phase of String theory the later being spontaneously
broken phase of the former.

There are two distinct approaches for the formulation of massless
Higher Spin gauge theory. One is a ``frame like `` formulation
\cite{Fradkin:1986qy} -- \cite{Vasiliev:1990en} and the other is a
``metric like `` formulation.\cite{Fronsdal:1978rb}
--\cite{Fronsdal:1978vb} (see also \cite{Francia:2002aa} for a
nonlocal formulation of massless Higher Spin fields on flat
space--time background,
 \cite{Francia:2005bu} for a
corresponding local formulation and \cite{Didenko:2003aa} for a
description of Higher Spin fields  on a `tensorial extension` of
AdS space). The problem of self consistent interactions of
massless Higher Spin fields has been effectively solved in the
``frame like`` formulation (see also \cite{Sezgin:2001zs}), where
it has been shown that
 the consistency of interaction requires the Anti-de Sitter (AdS) background.
 In order to address this problem in the ``metric like `` formulation
 we would like to discuss free Lagrangian and equations of motions first.
 Until now there are several descriptions available. Apart from the original construction
 of \cite{Fronsdal:1978vb} which involves off shell constraint on
 basic fields and gauge transformation parameters, and describes single irreducible Higher Spin
 mode one can obtain a Lagrangian description where all the fields and the gauge
parameters are
 unconstrained \cite{Buchbinder:2001bs} for irreducible  and \cite{Sagnotti:2003qa} for reducible
 Higher Spin fields on AdS.
 These descriptions include along with a basic field some auxiliary fields as well
and after a partial gauge fixing the Lagrangian of
\cite{Buchbinder:2001bs} reduces to the one of
\cite{Fronsdal:1978vb}. In the present paper we give an
alternative formulation to \cite{Sagnotti:2003qa} which, though
describes the same physical polarizations, is new in a sense that
it is carried out in a $({\cal D}+1)$-dimensional ambient space of
a ${\cal D}$- dimensional AdS. The field equations for irreducible
Higher Spin fields in an ambient space has been extensively
studied in \cite{Metsaev:1994ys}. Below we use this formalism for
Lagrangian description of reducible Higher Spin fields in the
framework of BRST approach. We believe that this formulation might
be useful for further studies of Higher Spin fields on AdS
background.

The plan of the paper is as follows. In Section \ref{FFTHS} we
review the free equations, the Lagrangian and the gauge
transformations which describe the reducible representations of
the Poincare group and of the Anti-de Sitter group, using the
triplet method for the description of Higher Spin fields. First we
deal with the flat space-time background. Then for the sake of
completeness
 we give some
general definitions and a review of various basic facts about
fields on ${\cal D}$ dimensional AdS space in terms of
representations of AdS isometry group. Then in Section \ref{TPAM}
we present a Lagrangian formulation of massless reducible bosonic
Higher Spin fields using an ambient space formalism when the
${\cal D}$ dimensional AdS space is embedded in a ${\cal D}+1$
dimensional flat
 {\it ambient space}. This formalism is further
 illustrated through the simple example
 of a vector field propagating through
 ${\cal D}$ dimensional AdS space. In Subsection \ref{MFAdS} we make
 some qualitative remarks on massive fields on AdS. In
the Section \ref{conl}, we present our conclusions and outlook. We
collect some useful formulas for calculations in ambient space in
the Appendix \ref{Ap}.

\section{Free Field Theory of Higher Spins}\label{FFTHS}

In this section we will review the construction of Higher Spin
field theories for massless (and massive) gauge fields.
 As it will become clear from what follows,
there is a particularly simple description of Higher Spin fields
 based
on the triplet construction, see \cite{Francia:2002pt}. We deal
exclusively with the case of fully symmetric Higher Spin tensors
as we stress in several points in what follows.

We now review the system described in \cite{Francia:2002pt} for
the case of a flat space - time and its AdS deformation  given in
\cite{Sagnotti:2003qa}, \cite{Barnich:2005bn}. This system is
named bosonic triplet and describes the propagation of reducible
representations of the Poincare and Anti de-- Sitter groups.
 The name
``triplet`` comes about, because a gauge invariant description of
massless fields with spins $s, s-2, s-4 ,...1/0$ requires in
addition, to a tensor field $\phi$ of rank $s$, the presence of
two more tensor fields. We denote them as $C$ (of rank $s-1$) and
$D$ (of rank $s-2$). After complete gauge fixing one is left only
with physical polarization of Higher Spin fields with spins
$s,s-2,s-4,..0$ or $1$ depending if $s$ is even or odd.

\subsection{Massless Fields on Flat
Space--Time.}\label{TPF}

 It is rather instructive,  to see how the
triplet formulation works for the simplest non--trivial example.
Consider a field $\phi_{\mu \nu} (x)$ of rank $2$. Therefore in
the triplet there should exist a field $C_\mu(x)$ of rank $1$ and
a field $D(x)$ of rank zero. The triplet equations take a rather
simple form in this case \cite{Francia:2002pt}:
\begin{equation}\label{graveqn}
\Box \phi_{\mu \nu} = \partial_\mu C_\nu + \partial_\nu C_\mu
\end{equation}
\begin{equation}
 C_{\mu} = \partial_\nu \phi_{\mu}{}^{\nu} - \partial _\mu D
\end{equation}

\begin{equation}
\Box D = \partial_\mu C^\mu.
\end{equation}
The system is invariant under the gauge transformations
\begin{equation}
\delta \phi_{\mu \nu} = \partial_\mu \lambda_\nu +
\partial_\nu \lambda_\mu,  \quad \delta C_\mu = \Box \lambda_\mu,
\quad \delta D = \partial_\mu \lambda^\mu.
\end{equation}
Let us introduce also a traceless field $\tilde \phi_{\mu \nu}$
\begin{equation}
\tilde \phi_{\mu \nu} = \phi_{\mu \nu} - \frac{1}{{\cal D}}
\eta_{\mu \nu}\phi^{'}, \quad \phi^{'} = \eta^{\mu \nu} \phi_{\mu
\nu}.
\end{equation}
In order to see the physical polarizations described by these
equations one can use the light-cone gauge fixing procedure i.e.,
eliminate $\tilde \phi_{++},\tilde \phi_{+i}$ and $\tilde
\phi_{+-}$  of the field $\tilde \phi_{\mu \nu}$ using gauge
transformation parameter $\lambda_{\mu}$. The other nonphysical
polarizations $\tilde \phi_{--},\tilde \phi_{-i}$ as well as the
field $C_\mu$ are eliminated by the field equations. Therefore one
is left with the physical degrees of freedom $\tilde \phi_{ij}$
$(i,j, = 1,..., {\cal D} -2)$ which correspond to the spin $2$
field and a gauge invariant scalar
\begin{equation}\label{scalar}
 \tilde D = \phi^{'} -2D.
\end{equation}
The system described above can be generalized to the case of
arbitrary spin as well. The field equations\footnote{Here the
total symmetrization with respect to the indexes is assumed. The
symbol $\partial \cdot$ means divergence, while $\partial$  is
symmetrized action of $\partial_\mu$ on a tensor as for example in
the r.h.s. of the first of \ref{graveqn}).} turn out to be
\begin{eqnarray}
&& \Box \; \phi \ = \ \partial C  \nonumber \ , \\
&& C = \partial \cdot \phi - \partial D \nonumber \ , \\
&& \Box \; D \ = \ \partial \cdot C \  \ . \label{triplet}
\end{eqnarray}
along with the gauge transformations
\begin{equation}\label{deltaphi}
\delta \phi = \partial \lambda, \quad \delta C = \Box \lambda,
\quad \delta D = \partial \cdot \lambda,
\end{equation}
~with  ~a ~completely ~unconstrained ~parameter ~of ~gauge
~transformations ~describe ~Higher ~Spin ~fields with spins $s,
s-2, ..1/0$.

The field equations in (\ref{triplet}) can be obtained as
equations of motion from a Lagrangian
\begin{equation}  {\cal L} =
\frac{1}{2}\ \phi \Box \phi \ + \ s\, \partial \cdot \phi \, C \ +
\ s(s-1)\,
\partial \cdot C \, D \
 - \ \frac{s(s-1)}{2} \, D \Box D \ - \ \frac{s}{2} \,
C^2,\label{Ltripletflat}
\end{equation}
which as it is clear from the discussion above contains except
from physical fields, some auxiliary fields as well. Let us note
that the case of spin two is somewhat degenerate, since only in
this case the combination (\ref{scalar}) is gauge invariant. For
the Higher Spin case we have
\begin{equation}
\delta (\phi^{'} -2D) = \partial \lambda^{'}.
\end{equation}
If one adds to the triplet equations ``by hand`` an extra
condition
 \be  \label{extra}
\phi^{'} = 2D, \ee then the parameter of the gauge transformations
$\lambda$ is no more unconstrained, but rather it obeys  the
vanishing trace condition
\begin{equation}
\lambda^{'}=0
\end{equation}
Further combining  equations (\ref{triplet}) and equation
(\ref{extra}) one obtains the condition
\begin{equation}
\phi^{''}=0.
\end{equation}
By $\phi^{'}$ we denote the trace $\phi ^\mu_{\mu \dots}$ and
$\phi^{''}$ the double trace $\phi^{\mu \nu}_{\mu \nu \dots}$.
 The extra condition (\ref{extra}) gives the possibility to
gauge away the fields  $C$, $ D$ and $\phi^{'}$ and thus one
 obtains the propagation of a single irreducible Higher Spin mode.
Alternatively using equations (\ref{triplet}) and (\ref{extra})
one can express fields $C$ and $D$ via field $\phi$ and put these
expressions back into the Lagrangian
 (\ref{Ltripletflat}) to obtain the one given by Fronsdal
\cite{Fronsdal:1978rb} which describes the propagation of a single
massless Higher Spin field. Let us note however that one can
obtain the equation (\ref{extra}) as an equation of motion from a
Lagrangian which contains some more auxiliary fields and therefore
formulate Lagrangians for Irreducible Massless Higher Spin fields
(\cite{Pashnev:1998ti}, \cite{Sagnotti:2003qa} for bosons, and
\cite{Buchbinder:2004gp} for fermions) without any off-shell
constraints on gauge transformation parameter and basic field
$\phi$.

\subsection{Fields on  AdS}\label{FAdS}

 Here we give some basic definitions concerning ${\cal D}$
dimensional Anti de Sitter space. More detailed treatment can be
found in \cite{Brink:2000ag} or in reviews \cite{deWit:2002vz}.

AdS space is a vacuum solution of Einstein equations with a
negative cosmological constant. Its Riemann tensor has a form
\begin{equation} R_{\mu\nu\rho\sigma} \ = \
\frac{1}{L^2} \, \left( g_{\mu\rho} \, g_{\nu\sigma} \ - \
g_{\nu\rho} \, g_{\mu\sigma} \right) \  \label{RADS},
\end{equation} where $L$ is an AdS radius and $L \rightarrow \infty$
corresponds to  a flat space-time limit. It is convenient to
represent   ${\cal D}$ dimensional AdS space with coordinates
$x^\mu$ ($\mu = 0.,.. ,{\cal D} -1$)and signature $(1,{\cal D}-1)$
as a hyperboloid in ${\cal D}+1$ dimensional flat space with the
signature $(2,{\cal D}-1)$, parameterized  by coordinates $y^A$
($A = 0.,.. ,{\cal D} $). The coordinates in this ambient space
obey the condition

\begin{equation}
\eta_{AB}y^A y^B = - L^2, \quad \eta_{AB} \eta^{AB} = {\cal D}+1,
\quad \eta_{AB} =(-,+,+ ... +,-).
\end{equation}
Therefore an isometry group is a pseudo-orthogonal group of
rotations $SO({\cal D}-1,2)$ and the AdS space itself is
isomorphic to a coset $SO({\cal D}-1,2)/SO({\cal D}-1,1)$.
 In order to simplify the
equations we set a radius of the AdS space to unity.

The AdS isometry group is noncompact and therefore its  unitary
representations are infinite dimensional. In order to build them
it is convenient to rewrite the $SO({\cal D}-1,2)$ algebra
\begin{equation}
[J_{AB}, J_{CD}]=\eta_{BC}J_{AD} -\eta_{AC}J_{BD} -
\eta_{BD}J_{AC}+ \eta_{AD}J_{BC},
\end{equation}
$$
 \quad J_{AB} = - J_{BA}, \quad
{(J_{AB})}^+ = -J_{AB},
$$
in a different form. Namely after taking the following linear
combinations
 \begin{equation} J_a^{\pm} =
(-iJ_{0a} \pm  J_{{\cal D}a}), \quad a= 1,...,{\cal D}-1
\end{equation}
 \begin{equation} H= iJ_{0{\cal D}},
\end{equation}  one obtains the commutation relations
 \bea
\left[H, J_a^{\pm}\right] = \pm J_a^{\pm} \nonumber \\
\left[J_a^-, J_b^{+}\right] = 2(H\delta_{ab} + J_{ab}) \nonumber \\
\left[J_{ab}, J_{\pm c}\right] = \delta_{bc}J_{\pm a} -
\delta_{ac}J_{\pm b}. \eea as well as
\begin{equation}
[J_{ab}, J_{cd}]=\eta_{bc}J_{ad} -\eta_{ac}J_{bd} -
\eta_{bd}J_{ac}+ \eta_{ad}J_{bc},
\end{equation}
>From these commutation relations one can conclude that AdS
isometry group has  a maximal compact subgroup
  $SO(2) \otimes
SO({\cal D}-1)$,
 spanned by generators $H$ and $J_{ab}$ respectively.
These  operators correspond to one dimensional and ${\cal D}-1$
dimensional  rotations. The operator  $H$ is an operator of energy
on AdS while time on AdS is defined as a variable  conjugate to
$H$. Therefore  time variable is  compact and  energy eigenvalues
are  quantized having integer values in order a wave function to
be single valued. However  usually one considers a
 covering space
of AdS, where time is uncompactified. The quadratic Casimir
operator in this basis has the from
\begin{equation} \label{Casimir}
{\cal C}_2 = - \frac{1}{2}J^{AB}J_{AB}= H(H - {\cal D} +1) -
\frac{1}{2}J^{ab}J_{ab} - J_a^+J_a^-
\end{equation}
Infinite dimensional unitary representations of the AdS group are
obtained  from the ``lowest weight states" $|E_0, s\rangle $,
which is a representation of $SO(2) \otimes SO({\cal D}-1)$. The
later therefore is
 characterized by an eigenvalue of energy  and by a Young tableaux
$s=(s_1,s_2,..,s_k)$, $k = [\frac{{\cal D} -1 }{2}])$. A lowest
weight state is annihilated by all operators $J_a^-$ \bea
J_a^-|E_0, s\rangle =0. \eea Then the other states of the
representations are obtained by successively applying operators
$J_a^+$ on a lowest weight state
\begin{equation}
J_{a_1}^+ J_{a_2}^+ ... J_{a_k}^+|E_0, s\rangle
\end{equation}
The crucial point is that representations obtained in this way do
not always have positive norm. Therefore, when building new states
with the help of operators $J_a^+$ one has to check their norm.
For some special values of $E_0$ and $s$ the norm is equal to
zero. There is a unitarity bound on the energy $E_0$ below which
the states get negative norms and should be excluded form the
physical spectrum. The unitarity bound is saturated (norm of
states becomes zero) for states with $E_0$ and $s$ related via
\begin{equation}
E_0 = s_1 + {\cal D} - t_1 -2
\end{equation}
where $t_1$ is the number of rows of maximal length $s_1$ in the
corresponding Young tableaux.\footnote{There might be some extra
states which saturate the unitarity bound. For example in the case
of ${\cal D} =4$ there are two states for scalar massless fields
with $E_0=1$ and $E_0=2$. These states have the same quadratic
Casimir operator but correspond to different asymptotic behavior
on the AdS boundary \cite{Fronsdal:1974ew}}. The states which
saturate the unitarity bound are identified with massless fields
on AdS space. These states decouple from the original multiplet
along with their descendants since their scalar product with the
other states is zero.
 This effect is known as a `multiplet shortening``
 and it is interpreted as an enhancement of gauge symmetry.

Fields whose energy is  above the unitarity bound are massive
representations of AdS space. Both massive and massless fields on
AdS have flat space -- time counterparts i.e., one can take an
usual flat space limit to obtain massless and massive fields
propagating through Minkowski space-time. However there is one
more type of fields on AdS, which have no flat space -time
analogue. These are called singletons. For example the unitarity
bound for spinless singleton is $E_0 = \frac{1}{2}({\cal D} -3).$
Singletons do not admit a proper field theoretical description in
AdS bulk, rather they are described as boundary degrees on
freedom.

Let us turn to a field theoretical description of massless fields
on AdS . In order to obtain  wave equations describing massless
fields with an arbitrary integer value of spin on AdS background
one has to find a relation between the quadratic Casimir operator
and the D'Alembertian. The result for totally symmetric
representations of AdS group i.e., when $s=(s,0,..0)$, is
\cite{Metsaev:1994ys}
  \be \label{uba} (\nabla^2 - (s-2)(s+{\cal D}-3))
F_{A_1,A_2...,A_s}(y)=0. \ee where
\begin{equation}\nabla^A =
\theta^{AB}\frac{\partial}{\partial y^B}, \quad \theta^{AB} =
\eta^{AB}+y^A y^B, \quad \nabla^2= \nabla^A \nabla_A.
\end{equation} A possible way to see where does
the condition  (\ref{uba}) come from is to introduce a auxiliary
Fock space spanned by a set of oscillators \be\label{osci} [
\alpha^A , \alpha^{B+} ] \ = \ \eta^{AB} \ ,   \ee and consider a
state in this Fock space
\begin{equation}
|\Phi \rangle\ = \frac{1}{s!}F_{A_1,A_2...,A_s} \alpha^{A_1 +}
\alpha^{A_2+}... \alpha^{A_s + }|0 \rangle.
\end{equation}
The generators of  $SO(2,{\cal D}-1)$ can be represented as
\begin{equation} \label{JAB}
J^{AB} = L^{AB}+M^{AB},
\end{equation}
where orbital part $L^{AB}$ and spin part $M^{AB}$ have the form
\begin{equation} \label{LAB}
L^{AB} = y^A \nabla^B - y^B \nabla^A, \quad M^{AB} = \alpha^{A+}
\alpha^B - \alpha^{B+} \alpha^A.
\end{equation}
A field $|\Phi \rangle$ in this Fock space is required to satisfy
the mass--shell condition
\begin{equation} \label{msh}
(\nabla^2 - m^2)|\Phi \rangle\ =0,
\end{equation}
where $m^2$ is  a ``mass -- like`` parameter  to be determined,
divergencelessness condition
\begin{equation}
\alpha^A \nabla_A|\Phi \rangle\ =0,
\end{equation}
 and transversality condition
\begin{equation}
y^A \alpha_A|\Phi \rangle\ =0.
\end{equation}
The requirement of invariance of these equations under gauge
transformations
\begin{equation} \delta |\Phi \rangle\ =\alpha^{A+}
\nabla_A|\Lambda_1 \rangle\ +y^A \alpha^+_A|\Lambda_2 \rangle\
\end{equation}
leads to the mass--shell  equation (\ref{uba}). If one computes
the explicit form of quadratic Casimir operator in terms of
realization (\ref{JAB})--(\ref{LAB}), one finds its eigenvalues
$<{\cal C}_2>$. Comparing equation
\begin{equation}
 ({\cal C}_2 -<{\cal C}_2>)|\Phi \rangle =0
 \end{equation}
 with (\ref{uba}) one obtains the expression for the unitarity
 bound in an alternative way.

\section{Triplet in  Ambient Space Formulation of AdS}\label{TPAM}

In order to make symmetries of Lagrangian and equations of motion
discussed in the previous subsection more transparent, we
reformulate them in the ambient space of $\cal D$ dimensional Anti
de--Sitter space. The system of free irreducible massless Higher
Spin fields propagating on an arbitrary dimensional Anti de --
Sitter space in the formalism of the ambient space was considered
in \cite{Fronsdal:1978vb} at Lagrangian level  and in
\cite{Metsaev:1994ys}
 at the level of equations of motion. Below we consider the case of
 reducible representations of AdS group
Let us note also another approach  \cite{Biswas:2002nk} where  a
dimensional reduction have been performed
 from flat
${\cal D} +1$ dimensional space to $\cal D$ dimensional Anti de
--Sitter space. Alternatively  we perform our construction  on the
Anti de -- Sitter space from the very beginning.

The isometry group of AdS space i.e., $SO(2,{\cal D}-1)$ can be
explicitly realized now in the whole construction.
 Further for simplicity
one can specify  the embedding by an extra  condition
\cite{Fronsdal:1974ew}
 \be \label{eco}
 y^A \frac{\partial}{\partial y^A} x^\mu =0
\ee i.e., choose  the $x$ -- space coordinates to be homogeneous
functions of degree zero in the ambient space coordinates. The
field's transformations from one coordinate system to another are
  \be
F_{\mu_1, \mu_2 ,... , \mu_s} (x) = \frac{\partial
y^{A_1}}{\partial x^{\mu_1}} \frac{\partial y^{A_2}}{\partial
x^{\mu_2}}... \frac{\partial y^{A_{s}}}{\partial x^{\mu_{s}}}
F_{A_1, A_2,...,A_s}(y).\label{xtoy} \ee Using this equation and
(\ref{thetadelta}) it is simple to find the inverse transformation
\be \frac{\partial x^{\mu_1}}{\partial y^{A_1}} \frac{\partial
x^{\mu_2}}{\partial y^{A_2}} ...\frac{\partial
x^{\mu_{s}}}{\partial y^{A_{s}}}
 F_{\mu_1 \mu_2,..., \mu_s}(x) =
\theta_{A_1}^{B_1} \theta_{A_2}^{B_2}...\theta_{A_s}^{B_s}
F_{B_1,B_2,...,B_s}(y). \ee

 Now we are in a position to reformulate the triplet equations in
 ambient space.
 It is convenient to take  a state in a Fock space in the form
\be | \tilde \Phi\rangle = \frac{1}{s!}\,  \tilde \Phi_{A_1 \ldots
A_s}(y) {\tilde\alpha}^{A_1+} \ldots \;{\tilde\alpha}^{A_s+}
|0\rangle, \ee where the oscillators are defined by the
transformation \be {\tilde\alpha}^{A}= \theta^A_B \alpha^B \ee and
therefore one has a commutation relation \be [{\tilde
\alpha}^{A},{\tilde\alpha}^{B+}]= \theta^{AB} . \ee
 Let us note that a state in ambient space satisfies identically
  \be \label{gk} T|\tilde \Phi \rangle =0, \quad T= y^A
\alpha_A.\ee This means that (\ref{gk}) does not impose any
further condition on $\tilde \Phi_{A_1 \dots A_s}$.

In order to describe a triplet on $\cal D$ dimensional Anti
de--Sitter space it is convenient to introduce the set of
oscillators $(\alpha^{\mu +},\alpha^\mu)$, which can be obtained
from the previous ones (\ref{osci}) and
 the AdS vielbein
 \begin{equation} [ \alpha^\mu , \alpha^{\nu +} ] \ = \ g^{\mu\nu},  \quad \alpha^\mu =e^\mu_a
 \alpha^a
\ , \end{equation}
 where $g_{\mu \nu}$ denotes the AdS metric.
An ordinary partial derivative is replaced by an operator
\begin{equation} p_\mu \ = \  -\; i \, \left(
\partial_\mu + \omega_{\mu}^{ab} \, \alpha_{\; a}^+\,
 \alpha_{ \; b} \right) \ . \label{covcurve}
\end{equation}
Acting with $p_\mu$ on a state in Fock space
\begin{equation}\label{PHI} |\Phi\rangle \ =\
\frac{1}{(s)!}\, \varphi_{\mu_1\mu_2...\mu_{s}}(x) \,
\alpha^{\mu_1 +} \ldots \alpha^{\mu_s +}  \, |0\rangle \ ,
\end{equation}
produces the proper covariant derivative
\begin{equation}
p_\mu|\Phi\rangle \ =-\frac{i}{(s)!}\alpha^{\mu_1 +} \ldots
\alpha^{\mu_s +} \nabla_\mu \, \varphi_{\mu_1\mu_2...\mu_{s}}(x)
|0\rangle,
\end{equation}
 where in (\ref{covcurve})   $\omega_\mu^{ab}$ denotes the spin --
connection on AdS and $\nabla_\mu$ is the AdS covariant
derivative. These operators satisfy commutation relations
  \begin{equation} [p_\mu,p_\nu] \
=  \alpha_{\; \mu}^+ \, \alpha_{\; \nu} \, -\, \alpha_{\; \nu}^+
\, \alpha_{\; \mu} \ , \end{equation} due to the  expression
(\ref{RADS}) for the Riemann  tensor.

Further let us introduce the following operators

 D'Alembertian
operator
\begin{equation} \label{lapl} l_0 \ = \
g^{\mu \nu} \, ( p_\mu p_\nu \, + \, i\; \Gamma^\lambda_{\mu
\nu}\; p_\lambda )\ = \ p^a \; p_a \, - \ i \, \omega_a{}^{ab} \,
p_b \end{equation} which  acts on Fock-space states as the proper
D'Alembertian operator
\begin{equation} l_0|\Phi\rangle \
=-\frac{1}{(s)!} \alpha^{\mu_1 +} \ldots \alpha^{\mu_s +}
 \Box \, \varphi_{\mu_1\mu_2...\mu_{s}}(x) \,
  \, |0\rangle \
\end{equation}

Divergence operator
\begin{equation}
l =  \alpha^\mu p_\mu \label{divergence}
\end{equation}
which acts on a state in the Fock space as divergence
\begin{equation} l|\Phi\rangle \
=-\frac{i}{(s-1)!}\alpha^{\mu_2 +} \ldots \alpha^{\mu_s +}
 \nabla_{\mu_1} \,
\varphi^{\mu_1}{}_{\mu_2\mu_3...\mu_{s}}(x) \,  \, |0\rangle \
\end{equation}

 Symmetrized exterior derivative operator,
\begin{equation}
l^+ =  \alpha^{\mu +} p_\mu \label{exteriord}
\end{equation}

\begin{equation} l^+|\Phi\rangle \
=-\frac{i}{(s+1)!} \alpha^{\mu +}\alpha^{\mu_1 +} \ldots
\alpha^{\mu_s +} \nabla_{\mu} \,
\varphi_{\mu_1\mu_2\mu_3...\mu_{s}}(x) \,   \, |0\rangle \
\end{equation}
which is hermitian conjugate to the operator $l$ with respect to
the scalar product
\begin{equation}
\int d^{\cal D} x \sqrt{-g} \langle \Phi_1||\Phi_2 \rangle .
\end{equation}
 It is straightforward to obtain the commutation relations of the
  algebra generated by these operators.
Having this algebra at hand one can construct the corresponding
nilpotent BRST charge. Then if one builds the nilpotent BRST
charge for this nonlinear algebra one arrives to a Lagrangian
description of a triplet on AdS according to the lines of
\cite{Sagnotti:2003qa}.

In order to extend the aforementioned construction in the ambient
space formulation, we should construct the ambient space operators
which are the analogues of $l, l^+, l_0$. It is straightforward to
obtain the following relations between various kinds of
derivatives in $x$ and $y$ spaces
\begin{equation}
\nabla^\mu \tilde \Phi_{\mu, \mu_1,...,\mu_s} =\frac{\partial
y^{A_1}}{\partial x^{\mu_1}}... \frac{\partial y^{A_s}}{\partial
x^{\mu_s}}(\nabla^A + ({\cal D} + s)y^A) \tilde
\Phi_{A,A_1,...,A_s}, \label{tran1}
\end{equation}
\begin{equation}
\nabla_{(\mu_1} \tilde \Phi_{ \mu_2,...,\mu_s)} =\frac{\partial
y^{A_1}}{\partial x^{\mu_1}}... \frac{\partial y^{A_s}}{\partial
x^{\mu_s}}(\partial_{(A_1}\Phi_{A_2,...,A_s)}
 + (s-1)y^A \eta_{(A_1 A_2} \tilde \Phi_{A,A_3,...,A_s)}),
\label{tran2}
\end{equation}
 \begin{eqnarray} \nonumber
\Box \tilde \Phi_{\mu_1... \mu_s} &=& \frac{\partial
y^{A_1}}{\partial x^{\mu_1}}... \frac{\partial y^{A_s}}{\partial
x^{\mu_s}}( \nabla^2 \Phi_{A_1...A_s}   - s\Phi_{A_1... A_s} +2s
\partial_{(A_1} y^A\Phi_{A,A_2...A_s)}\\
&& + s(s-1)y^A y^B\eta_{(A_1 A_2}\Phi_{A,BA_3...A_s)} ),
\end{eqnarray}
\be \Box \tilde \Phi_{A_1, ... A_s} = \nabla^2 \tilde
\Phi_{A_1,...A_s}. \ee
 Right
hand sides of equations (\ref{tran1}) and (\ref{tran2}) can be
obtained acting on Fock space states by operators \be e = i{\tilde
\alpha^A} \nabla_A + i({\cal D} + {\tilde \alpha^{A +}} {\tilde
\alpha _A})T - i T^+{\tilde \alpha^{A }} {\tilde \alpha _A} ,
 \ee
\be e^+ =i{\tilde \alpha^{A+}} \nabla_A -i T^+{\tilde \alpha^{A
+}} {\tilde \alpha _A} + i {\tilde \alpha^{A+ }} {\tilde
\alpha^+_A} T, \ee respectively.  Then the operator $e_0$ is by
definition the one obtained by their commutation
 \be \label{e0}
e_0= [ e, e^+ ], \ee and its explicit form is
\begin{eqnarray}
e_0&=& -\nabla^2  +2 (\alpha^{A +} y_A) (\nabla_B {\tilde
\alpha^B}) - 2 ({\tilde \alpha^{A^+}} \nabla_A) (\alpha^B y_B)  \\
\nonumber &&+{\cal D} {\tilde \alpha}^{A +}{\tilde \alpha}_{A}  -
{\tilde \alpha^{A +}}{\tilde \alpha_{A}} + {({\tilde \alpha}^{A +}
{\tilde \alpha}_{A})}^2 - ({\tilde \alpha^{A +}}{\tilde
\alpha_{A}^+})({\tilde \alpha^{B }}{\tilde \alpha_{B}}) \\
\nonumber &&+2 T^+({\tilde \alpha}^{A +}{\tilde \alpha}_{A})T
  -({\tilde \alpha^{A +}}{\tilde \alpha_{A}^+})TT -T^+T^+({\tilde
\alpha^{A }}{\tilde \alpha_{A}}) - {\cal D} T^+T
\end{eqnarray}
 The remaining commutation relations look as follows \be [ e_0, e] = 2(1-{\cal D}) e - 4
\alpha^{A +} \alpha _A e +4 e^+ ({\tilde \alpha}^{A }{\tilde
\alpha_{A}}). \ee Finally one   constructs the BRST charge for
this system
\begin{eqnarray} \label{brstAM}
Q& =&c_0( e_0  + 6 - 2 {\cal D} - 4{\tilde \alpha^{A +}} {\tilde
\alpha _A} )
   +  c e^{+} +  c^+  e
     -  c^+  c  b_0 \\ \nonumber
&&+ ( 2{\cal D} + 4{\tilde \alpha^{A +}} {\tilde \alpha _A} -6 )
c_0  b^+  c +
      ( 2{\cal D} + 4{\tilde \alpha^{A +}}{\tilde \alpha _A} -6 )  c_0  c^+
b \\ \nonumber &&-4 c_0 c^+b^+ ({\tilde \alpha^{A }}{\tilde
\alpha_{A}}) + 4 c_0 c b   ({\tilde \alpha^{A +}}{\tilde
\alpha_{A}^+})
      +12  c_0  c^+  b^+  c  b .
\end{eqnarray}
where we have introduced ghost variables $c_0, c^+, c$ with ghost
number $+1$ and corresponding antighost variables $b_0, b, b^+$
with ghost number $-1$. After defining the ghost vacuum as
\begin{equation}
b_0 | 0 \rangle_{gh.}=c | 0 \rangle_{gh.}=b | 0 \rangle_{gh.}=0
\end{equation}
one obtains a Lagrangian
\begin{equation} \label{lagrangian1}
{\cal L}= \int d c_0  \langle \tilde \Phi| Q |\tilde \Phi \rangle
\end{equation}
which is invariant under gauge transformations
\begin{equation} \label{gauge1}
\delta |\tilde \Phi \rangle = Q |\Lambda \rangle.
\end{equation}
The total vacuum is now a direct product of ghost and $\tilde
\alpha$ vacua
\begin{equation}
 |0 \rangle = |0 \rangle_{\tilde \alpha} \otimes |0 \rangle_{gh.},
 \quad \tilde \alpha^A|0 \rangle_{\tilde \alpha} =0
\end{equation}
and the integration of the ghost zero mode is defined as
\begin{equation}
\int d c_0  \langle 0| c_0 |0 \rangle=1.
\end{equation}
One can conclude from (\ref{lagrangian1}) and (\ref{gauge1}) that
in order for the Lagrangian to have ghost number zero the field
$|\tilde \Phi \rangle$ must have ghost number zero, while gauge
transformation parameter $|\Lambda \rangle$ must have ghost number
$-1$. Therefore their expansion in terms of ghost variables has
the form
\begin{equation}
|\tilde \Phi \rangle = |\phi \rangle + c^+ b^+|D \rangle  + c_0
b^+|C \rangle,
\end{equation}
\begin{equation}
|\Lambda \rangle =   b^+|\lambda \rangle
\end{equation}
where states $|\phi \rangle, |D \rangle, |C \rangle $ and
$|\lambda \rangle$ are expanded in terms of oscillators $\tilde
\alpha^{A+}$.

Using relations above it is straightforward to obtain the
corresponding Lagrangian
\begin{eqnarray}\label{LAdSAM}
{\cal L}& =&  \langle \phi | e_0  + 6 - 2 {\cal D} - 4{\tilde
\alpha}^{A +}{\tilde \alpha}_{A} | \phi \rangle - \langle D | e_0
+6 + 2 {\cal D} + 4 {\tilde \alpha}^{A+} {\tilde \alpha}_A |
D \rangle + \langle C|| C \rangle   \nonumber \\
&& - \langle \phi | e^+ | C \rangle + \langle D | e  |C \rangle -
\langle C | e | \phi \rangle + \langle C| e^+ | D \rangle
\\ \nonumber &&+4 \langle D |   {\tilde \alpha}^{A }{\tilde
\alpha}_{A} | \phi \rangle +4\langle \phi | {\tilde
\alpha}^{A+}{\tilde \alpha}^+_{A} | D \rangle,\end{eqnarray} gauge
transformations \be \label{GTAM} \delta |\phi  \rangle = e^+
|\Lambda  \rangle, \quad \delta |C \rangle = e_0 |\Lambda
\rangle, \quad \delta |D \rangle = e |\Lambda  \rangle, \ee and
equations of motion
\begin{eqnarray}\label{EQNAM}
&& (e_0 + 6 - 2 {\cal D} - 4{\tilde \alpha}^{A +}{\tilde
\alpha_{A}} )| \phi \rangle= e^+  | C \rangle
- 4{\tilde \alpha}^{A +}{\tilde \alpha}_{A}^+|D \rangle  , \\
&&  | C \rangle =  e  |\phi \rangle  -  e^+ | D \rangle \ , \\
&& (e_0 +6 + 2 {\cal D} + 4 {\tilde \alpha}^{A+} {\tilde \alpha}_A
)  |D \rangle= e  | C \rangle + 4 {\tilde \alpha^{A }}{\tilde
\alpha}_{A}  | \phi \rangle.
\end{eqnarray}
Equations (\ref{GTAM}) and  (\ref{EQNAM}),  can be easily written
in component notation using (\ref{tran1}, \ref{tran2}). Namely for
gauge transformations we have
\begin{eqnarray}\label{CGTAM}
&&\delta \phi= (s-1) \eta (y \cdot \Lambda) + \partial \Lambda,
\nonumber \\
&&\delta D= ({\cal D} +s -2)(y \cdot \Lambda) + \nabla \cdot
\Lambda, \\
&&\delta C= \nabla^2 \Lambda + 2(s-1) \partial(y \cdot \Lambda) +
(s-1)(s-2)\  \eta ((yy)\cdot \Lambda) \nonumber \\
&&+ (s-1)(2-s-{\cal D})\Lambda +2 \eta \Lambda', \nonumber
\end{eqnarray}
where as before $\partial$ means action of the derivative
$\partial^A$, $\nabla \cdot$  means divergence and a total
symmetrization with respect to the indices is  assumed. In
addition we have defined:
\begin{equation}\label{notation}
(y \cdot \phi)_{A_1, \dots A_{s-1}}= y^A \phi_{A A_1, \dots
A_{s-1}}, \ \ (\phi')_{A_1, \dots A_{s-2}}= \theta^{A B}
\phi_{A,B,A_1,\dots A_{s-2}}
\end{equation}
and
\begin{equation}
\eta \phi = \eta_{(A_1A_2}\phi_{A_3,\dots A_s) },
\end{equation}
while equations of motion look as follows:
\begin{eqnarray}\label{CEQNAM}
&& (\nabla^2-[(2-s)(3-{\cal D}-s)] ) \phi + 2s \partial (y\cdot
\phi) + s(s-1) \ \eta ((yy)\cdot \phi) + 2 \eta \phi'= \nonumber
\\
&&= (s-1) \ \eta (y \cdot C) + \partial C + 8 \eta D,\\
&& C= ({\cal D}+s-1) y\cdot \phi + \nabla \cdot \phi- (s-2) \
\eta(y\cdot D) - \partial D, \nonumber \\
&&(\nabla^2-[s({\cal D}+s-1)+4] ) D +2(s-2) \partial (y \cdot D)
+ \nonumber \\
&&(s-2)(s-3)\ \eta ((yy)\cdot D) + 2 \eta D' = ({\cal D}+s-2)
y\cdot C + \nabla \cdot C -4 \phi'.\nonumber
\end{eqnarray}
Again as it was for the case for a triplet in flat space time one
can add by hand an extra condition \be {\tilde \alpha}^{A }{\tilde
\alpha}_{A} |  \phi \rangle = 2| D \rangle \ee and restrict the
parameter of gauge transformations by the condition ${\tilde
\alpha}^{A }{\tilde \alpha}_{A} | \Lambda \rangle = 0$ to obtain
the  Lagrangian description \cite{Fronsdal:1978vb}
\begin{eqnarray}
{\cal L}&=& \langle \phi| e_0 - e^+e + \frac{1}{2} e^+ e^+ {\tilde
\alpha^{A }}{\tilde \alpha_{A}} + \frac{1}{2}{\tilde \alpha^{A
+}}{\tilde \alpha_{A}^+} e e - \frac{1}{2} {\tilde \alpha^{A
+}}{\tilde \alpha_{A}^+} e_0 {\tilde \alpha^{B }}{\tilde
\alpha_{B}} \\ \nonumber
 && - \frac{1}{4} {\tilde \alpha^{A +}}{\tilde \alpha_{A}^+}
   e^+ e
   {\tilde \alpha^{B }}{\tilde \alpha_{B}} +
6 - 2 {\cal D} - 4{\tilde \alpha^{A +}}{\tilde \alpha_{A}} \\
\nonumber &&- \frac{1}{2} {\tilde \alpha^{A +}}{\tilde
\alpha_{A}^+}
 (1+ {\cal D} +
2{\tilde \alpha^{B +}}{\tilde \alpha_{B}}) {\tilde \alpha^{C
}}{\tilde \alpha_{C}} |\phi \rangle
\end{eqnarray}
of a single irreducible Higher Spin mode on AdS space. The
integration measure for the action in ambient space is given in
\cite{Fronsdal:1974ew}
\begin{equation}\label{measure}
 d^{{\cal D} +1} y \delta^{{\cal D}+1}(y^2+1) =
d^{{\cal D}} x \sqrt{-g}
\end{equation}

\subsection{A Vector Field on AdS Background}\label{U1}
It will be useful to show explicitly through a specific example
how the formalism described in the previous subsection leads to a
description of massless fields on AdS background. For this we
consider the simplest case of a massless vector field
 on AdS background.
 The Lagrangian (\ref{LAdSAM}) which in this case contains only one
 physical field $\phi_A$ and an auxiliary field $C$
 is:
\begin{eqnarray} \nonumber
{\cal L} &=& \theta^{AB}\phi_A(\nabla^2 + ({\cal D} -2))\phi_B +2
\phi_A\nabla^A(y^B\phi_B)- C^2 \\
&& - (\nabla_A C ) \phi^A+ (\nabla_A \phi^A) C + {\cal D} C y^A
\phi_A. \label{AL}
\end{eqnarray}
 The equations of motion in
(\ref{CEQNAM}) become:
\begin{equation}
\nabla^2 \phi_A + ({\cal D} -2)\phi_A + 2 \partial_A(y^B \phi_B) =
\partial_A C, \label{AE1}
\end{equation}
\begin{equation}
\nabla_A \phi^A + {\cal D} y^A \phi_A = C, \label{AE2}
\end{equation}
with gauge transformation rules (\ref{CGTAM}):
\begin{equation}
\delta \phi_A = \partial_A \lambda, \quad \delta C = \nabla^2
 \lambda.
\end{equation}
One might suppose that when making an embedding into higher
dimensional space some extra degrees of freedom can appear, for
example a vector in an ambient space might correspond to a vector
and a scalar in $x$ space. It is not so however since in ${\cal
D}+1$ dimensions one has some extra gauge freedom which allows to
eliminate   ``extra" degrees of freedom. In particular one can
impose the gauge condition $y^A \phi_A=0$, which eliminates a
scalar from the spectrum.
  Then one is left
 with a gauge parameter $\tilde \lambda$ which is
 constrained through $y^A \partial_A {\tilde \lambda} =0$.
 Using this parameter one can further gauge away the field $C$ and
 check that this further gauge fixing procedure is consistent with
 equations of motion and therefore is correct  \cite{Metsaev:1994ys}.
 Finally one is left with equations
 \begin{equation}
 (\nabla^2 + {\cal D} -2) \phi_A=0, \quad \partial^A \phi_A = y^A \phi_A
 =0,
 \end{equation}
which  describe  a massless vector field on AdS. It is simple to
rewrite this system in the $x$-- space. Namely the Lagrangian
(\ref{AL}), which   contains physical field $\phi_\mu$ and an
auxiliary field  $C$, takes the form
\begin{equation}
{\cal L} =   \frac{1}{2} \phi^\mu \Box \phi_\mu+ (\nabla_\mu
\phi^\mu) C- \frac{1}{2} C^2 + \frac{1}{2}({\cal D} -1) \phi^\mu
\phi_\mu. \label{Maxl}
\end{equation}
This Lagrangian is invariant under gauge transformations
\begin{equation}\label{gauge} \delta \phi_\mu  = \nabla_\mu \lambda, \quad
\delta C = \Box \lambda
\end{equation}
and gives the following equations of motion
\begin{equation}
\Box \phi_\mu = \nabla_\mu C - ({\cal D} -1) \phi_\mu
\label{maxe1}
\end{equation}
\begin{equation}
\nabla_\mu \phi^\mu = C.
\end{equation}
We can use (\ref{gauge}) to gauge away C and then we get:
\begin{equation}\label{AdSU1}
(\Box + {\cal D} - 1) \phi_\mu=0, \quad \nabla^\mu \phi_\mu=0.
\end{equation}

\subsection{Massive Fields on AdS}\label{MFAdS}

Although it is beyond the main scope of this paper, we discuss
briefly how one could proceed in the case of massive fields in
AdS. This construction has been outlined in
\cite{Buchbinder:2005ua} where the detailed study of massive
Higher Spin fields on flat space was performed in the framework of
BRST approach. We briefly summarize their arguments.

The crucial point in this construction is played by the method of
auxiliary representations (see \cite{Burdik:2000kj} for a review).
This method can be more transparently demonstrated in the simpler
example for the $SO(2,1)$ algebra.

Suppose we want to build a BRST charge for the constraints $M=
\frac{1}{2} \tilde \alpha^A \tilde \alpha_A$, $M^+= \frac{1}{2}
\tilde \alpha^{A +} \tilde \alpha_A^+$,
 $N=  \tilde \alpha^{A +} \tilde \alpha_A + \frac{{\cal D}}{2}$.
 Though these operators form a closed algebra, the
operator $N$ cannot be included into the total set of constraints
since it is strictly positive. The way out, is to introduce an
extra oscillator $[b,b^+]=-1$ and build an auxiliary realization
of the algebra in terms of these oscillators and some parameter
$h$. The only requirement for the $M_{aux}, M^+_{aux}, N_{aux}$ is
that they obey the same commutation relations as the original
operators and that $N_{aux}$ depends on $h$ linearly. For the
particular example under consideration these representations can
be chosen  as
 \be M_{aux}= b \sqrt{h+1 + b^+b}, \quad M_{aux}^+ =
b^+ \sqrt{h+ b^+b}, \quad N_{aux} = - 2b^+b -h. \ee Next one
defines modified operators $ \tilde M,  \tilde M^+,  \tilde N$
 as the sum of old and new operators and build the BRST charge
\be Q = c_N \tilde N + c_M
 \tilde M^+ + c_M^+ \tilde M + c_{N}(2c_N^+ b_N + 2 b_N^+ c_N -2) - c_M^+
c_M b_N \ee for the system of modified constraints. Finally one
considers an auxiliary phase space $(x_h,h)$ such that $[x_h,h]=i$
and performs similarity transformation $Q \rightarrow U^{-1} Q U$
with $U= e^{i \pi x_h}$ and $\pi = N - 2b^+ b - 2 + 2c_M^+ b_M +
2b_M^+c_M$. In this way one eliminates the dependence of the BRST
charge $Q$ on the offending ghost variables $c_N $ and $b_N$ but
maintains its nilpotency at the same time.

This procedure is completely general in the sense that it can be
used for BRST construction of algebras either linear or nonlinear
when some of the Cartan generators (operator $N$ in our case) are
excluded from the set of constraints due to some physical reasons
(i.e., because it is strictly positive). The problem thus reduces
to the question of how to build  auxiliary representations for the
algebra under consideration and as a result a general state in
Fock space will depend also on extra oscillators $b^+_i$.

An analogous picture appears for the case of massive Higher Spin
fields \cite{Buchbinder:2005ua} since equation (\ref{e0}) becomes
now
\begin{equation} \label{l0m}
 [ e , e^+ ] \ = \tilde{e}_0 - m^2 \ ,
\end{equation}
where $\tilde e_0$ is now a ``modified D`Alembertian" for ${\it
massive}$ bosonic fields on AdS i.e., the mass parameter plays the
role of ``central charge`` in  the algebra. The auxiliary
representations for the nonlinear algebra under consideration was
built in \cite{Burdik:2002rz} in terms of two sets of oscillators
and two parameters which correspond to mass and spin. It is an
interesting open problem to carry out explicit calculations for
the Lagrangian describing massive Higher Spin fields on AdS, as
well as to study its supersymmetric extension and the mechanism of
mass generation especially in the framework of AdS/CFT duality
\cite{Zinoviev:2001dt} -- \cite{Bianchi:2005ze}.

\section{Conclusions}\label{conl}
In this paper, we have studied some aspects of Lagrangian
formulation for Higher Spin gauge fields in AdS$_{\cal D}$
background, by using the triplet method for describing Higher Spin
fields. We have embedded AdS$_{\cal D}$ into a ${\cal D} +1$
dimensional flat spacetime, the ambient space. Consequently we
generalized the AdS spacetime triplet formalism to its ambient
space analogue. We further demonstrated the equivalence of the two
formulations for the simple case of a U(1) gauge field on
AdS$_{\cal D}$.

The formulation in terms of ambient space might be useful when
considering self interaction of the triplet
\cite{Buchbinder:2006eq} on AdS background and for a further study
of the properties of massless and massive Higher Spin fields.

A further application might be in the study of a possible
connection between massless and massive Higher Spin theory with
superstring theory. Namely it is interesting to understand the
properties of string theory in the high energy limit
\cite{Moeller:2005ez}--\cite{Gross:1987ar} especially on a highly
curved Anti de Sitter background as well as a further study of
AdS/CFT duality beyond the Supergravity limit
\cite{Sezgin:2002rt}. One more topic of interest is to perform the
analogous study for mixed symmetry fields on AdS
\cite{Alkalaev:2006rw}, (see \cite{Burdik:2001hj} --
\cite{Bekaert:2006ix} for a recent discussion for flat space--time
in ``metric like`` approach).

\vspace{1cm}

\noindent {\bf Acknowledgments.} It is a pleasure to acknowledge
X. Bekaert, I. Buchbinder, N. Irges, A. Koshelev, P.~ Lavrov, A.
Petkou,
 D. Sorokin, P. Sundell,  T. Tomaras, M. Vasiliev,
 P. West and R.Woodard  for the useful discussions.
 The work of A.F. has been supported by a "Pythagoras" fellowship
 by the Greek Ministry of Education.  The work of A.F. and
 M.T. was supported by the European contract MRTN-CT-2004-512194.


\renewcommand{\thesection}{A}

\setcounter{equation}{0}

\renewcommand{\theequation}{A.\arabic{equation}}

\section{Some Formulas in Ambient Space}\label{Ap}

One can check some useful relations for an ambient space
 \be
\theta^{AB} \theta^{BC} = \theta^{AC},  \quad \nabla^C\theta^{AB}
=  \theta^{CA}y^B + \theta^{CB} y^A,  \quad \nabla^A\theta^{AB} =
{\cal D}y^B, \ee
 \begin{equation}
 [\nabla_A,\nabla_B] =    -y_A \nabla_B +y_B \nabla_A,
\ee
\begin{equation}
 [\nabla^2,y^A] = 2 \nabla_A +{\cal D} y^A,  \quad  [\nabla^2,\nabla^A] =
 (2 -{\cal D} )\nabla^A + 2 y^A \nabla^2.
\ee

\begin{equation}
 y^A \nabla_A =0,  \quad  y^A  \theta_A^B =0,  \quad  \nabla^A y_A = {\cal
D}. \ee The induced metric, its inverse and Christofell connection
look as follows: \be g_{\mu \nu} = (\partial_\mu y^A)
(\partial_\nu y^A),  \quad g^{\mu \nu} = (\nabla^A x^\mu)(\nabla^A
x^\nu), \quad \Gamma^\lambda_{\mu \nu}=
 \frac{\partial x^\lambda}{\partial y^A} \frac{\partial^2 y^A}{\partial
x^\mu \partial x^\nu}. \ee
 We have also
\be \label{thetadelta} \theta^A_B = \frac{\partial y^A}{\partial
x^\mu} \frac{\partial x^\mu}{\partial y^B}, \quad \delta^\mu_\nu =
\frac{\partial y^A}{\partial x^\nu} \frac{\partial x^\mu}{\partial
y^A}, \ee as well as \be \theta^{AB} = g^{\mu\nu} ({\partial_\mu
y^A}) ({\partial_\nu y^B}), \ee which follow from the
differentiation rules
 \be \theta^{AB}\frac{\partial }{\partial
y^B}
 = \eta^{AC} \frac{\partial x^\mu}{\partial y^C} \frac{\partial }{\partial
x^\mu},
 \quad
\frac{\partial }{\partial x^\mu}
 =  \frac{\partial y^A}{\partial x^\mu} \frac{\partial }{\partial
 y^A} \label{derivatives}.
\ee Note that
\begin{equation} \nabla^A x^\mu =  \frac{\partial
x^\mu}{\partial y^A }
\end{equation} due to the embedding
condition (\ref{eco}).
 Finally it is
straightforward to derive the following relations \be
\theta^{AB}\frac{\partial x^\mu}{\partial y^B} =
g^{\mu\nu}\frac{\partial y^A}{\partial x^\nu}, \quad
 \nabla^2  x^\mu = - \Gamma^\mu_{\nu \rho}g^{\nu \rho}.
 \end{equation}
\begin{equation}
\nabla^A \nabla_A \Phi^{C_1,C_2,,...C_s} = \nabla^\mu\nabla_\mu
\Phi^{C_1,C_2,,...C_s}, \quad
 {\nabla}_\mu \frac{\partial y^A}{ \partial
x^\nu}= g_{\mu \nu}y^A. \label{imptr}\ee

\end{document}